\documentclass[12pt,leqno]{article}
\usepackage{latexsym}
\usepackage{amsfonts}
\usepackage{amsmath}
\usepackage{amsthm}
\usepackage{hyperref}

\def\tensor{\otimes}
\def\d{\partial}

\def\id{\mathbf 1}
\def\3{\underline }
\def\5{\bar }
\def\6{\partial }
\def\7{\hat }
\def\4{\tilde }
\def\8{\bf }
\def\9{\dot }

\newcommand{\dl}[1]{\displaystyle\frac{{\d}}{\d #1}}

\newcommand{\dd}[2]{\frac{\d #1}{\d #2}}
\newcommand{\ddr}[2]{\frac{\d^R #1}{\d #2}}
\newcommand{\ddl}[2]{\frac{\d^L #2}{\d #1}}

\newcommand{\vddr}[2]{\frac{\delta^R #1}{\delta #2}}

\newcommand{\vddll}[2]{{\frac{\delta^L #1}{\delta #2}}}

\newcommand{\pb}[2]{\left\{{}#1{},{}#2{}\right\}}

\newcommand{\inner}[2]{\langle #1,\,#2\rangle}

\newcommand{\commut}[2]{[#1,#2]}

\def\half{{\frac{1}{2}}}
\newcommand{\gh}[1]{{\rm gh}(#1)}

\newcommand{\p}[1]{|#1|}

\newcommand{\ttime}{t}

\newcommand{\bref}[1]{{\rm {\bf\ref{#1}}}}

\def\be{\begin{eqnarray}}
\def\ee{\end{eqnarray}}
\def\beann{\begin{eqnarray*}}
\def\eeann{\end{eqnarray*}}
\def\beq{\begin{equation}}
\def\eeq{\end{equation}}
\def\ba{\begin{array}}
\def\ea{\end{array}}
\def\ben{\begin{enumerate}}
\def\een{\end{enumerate}}
\def\bea{\begin{eqnarray}}
\def\eea{\end{eqnarray}}
\def\beann{\begin{eqnarray*}}
\def\eeann{\end{eqnarray*}}
\def\beq{\begin{equation}}
\def\eeq{\end{equation}}
\def\ba{\begin{array}}
\def\ea{\end{array}}
\def\ben{\begin{enumerate}}
\def\een{\end{enumerate}}

\def\5{\bar }
\def\6{\partial }
\def\7{\hat }
\def\4{\tilde }

\def\cA{{\cal A}}
\def\cB{{\cal B}}

\def\cH{{\cal H}}

\def\cM{{\cal M}}

\def\cP{{\cal P}}

\def\manM{{\mathcal M}}

\def\s0#1#2{\mbox{\small{$\frac{#1}{#2}$}}}

\def\qed{\hbox{${\vcenter{\vbox{
\hrule height 0.4pt\hbox{\vrule width 0.4pt height 6pt
\kern5pt\vrule width 0.4pt}\hrule height 0.4pt}}}$}}

\newcommand{\densor}{\otimes}

\numberwithin{equation}{section}

\begin{document}

\begin{titlepage}

\begin{flushright}
ULB-TH/03-34\\
hep-th/0310083
\end{flushright}

\begin{centering}

\vspace{0.5cm}

{\bf{\Large Hamiltonian BRST and Batalin-Vilkovisky formalisms for
second quantization of gauge theories}}

\vspace{1.5cm}

{\large Glenn Barnich$^{*,a}$ and Maxim Grigoriev$^{\dag,a,b}$}

\vspace{1.5cm}

$^a$Physique Th\'eorique et Math\'ematique,\\ Universit\'e Libre
de Bruxelles,\\
Campus Plaine C.P. 231, B--1050 Bruxelles, Belgium

\vspace{.5cm}

$^b$Tamm Theory Department, Lebedev Physical Institute,\\
Leninsky prospect 53, 119991 Moscow, Russia

\vspace{1.5cm}

\end{centering}
\vspace{.5cm}

\begin{abstract}
Gauge theories that have been first quantized using the
Hamiltonian BRST operator formalism are described as classical
Hamiltonian BRST systems with a BRST charge of the form
$\langle \Psi,\hat \Omega\Psi\rangle_{\rm even}$ and with natural ghost and
parity degrees for all fields. The associated proper solution of
the classical Batalin-Vilkovisky master equation is constructed
from first principles. Both of these formulations can be used as
starting points for second quantization. In the case of time
reparametrization invariant systems, the relation to the standard
$\langle \Psi,\hat \Omega\Psi\rangle_{\rm odd}$ master action is established.
\end{abstract}

\vspace{1.5cm}

\noindent \footnotesize{$^*$ Research Associate of the National
Fund for
  Scientific Research (Belgium).\\
  $^\dag$ Postdoctoral Visitor of the National Fund for Scientific Research (Belgium).}

\end{titlepage}

\tableofcontents

\vfill


\section{Introduction}

It has been realized in
\cite{Thorn:1987qj,Bochicchio:1987zj,Bochicchio:1987bd} (see also
\cite{Thorn:1989hm}) that the action of open bosonic string field
theory \cite{Siegel:1988yz,Witten:1986cc}, with free part given by
the expectation value of the BRST operator, should be understood
as a solution to the classical Batalin-Vilkovisky (BV) master
equation. The collection of fields, ghosts, and all the ghosts for
ghosts corresponds to the coefficients of the states in negative
ghost numbers, while the associated antifields correspond to the
coefficients of the states in positive ghost numbers. In the
standard gauge field theory context, however, a proper solution to
the Batalin-Vilkovisky master equation is obtained from a gauge
invariant action, a generating set for its non trivial gauge
symmetries and, if needed, associated reducibility operators
\cite{Batalin:1981jr,Batalin:1983wj,Batalin:1983jr,Batalin:1984ss,%
Batalin:1985qj} (see also \cite{Henneaux:1992ig,Gomis:1995he} for
reviews).

The purpose of this paper is to construct from basic principles
the proper solution of the master equation associated to a theory
first quantized using the Hamiltonian BRST operator (or "BFV")
formalism \cite{Fradkin:1975cq,Batalin:1977pb,Fradkin:1978xi} (see
also \cite{Henneaux:1985kr}) and to relate it with the standard
master action of
\cite{Thorn:1987qj,Bochicchio:1987zj,Bochicchio:1987bd}. This
involves several steps:
\begin{enumerate}
\item the reformulation of BRST quantum mechanics as a classical Hamiltonian
BRST system;

\item using the known proper solution of the master equation for Hamiltonian BRST systems;

\item for time reparametrization invariant systems,
relating the constructed master action to the standard one by
showing that they differ by the quantization of classically
trivial pairs.

\end{enumerate}
The first two steps are treated in section \bref{sec:BFV-BV},
while section \bref{ri} is devoted to the last step.

More precisely, for the first step,
it has been pointed out by many authors (see e.g.~%
\cite{Kibble:1979tm,Heslot:1985xx} and
\cite{Hatfield:1992rz,Schilling:1996xx,Ashtekar:1997ud} for
reviews and further references) that the Hilbert space of quantum
mechanics can be understood as a (possibly infinitedimensional)
symplectic manifold and that the Schr\"odinger evolution appears
as a Hamiltonian flow on this phase space. This point of view
provides a useful set-up for second quantization. In order to
apply these ideas to gauge systems quantized in the operator
formalism according to the Hamiltonian BRST prescription, one also
needs to understand in this context the physical state condition
$\hat \Omega \psi=0$, as well as the identification of BRST closed
states up to BRST exact ones. The latter two problems alone have
been faced in the context of string field theory
\cite{Witten:1986cc,Siegel:1988yz,Thorn:1989hm,Zwiebach:1993ie},
with the somewhat surprising conclusion that the object
$\langle \Psi,\hat\Omega\Psi\rangle$ is not a BRST charge, but a solution to
the master equation. This is due to the fact that the ghost pair
associated to the mass shell constraint is quantized in the
Schr\"odinger representation.

In our approach, we will start by assuming that the number of
independent constraints is even so that there is also no
fractionalization of the ghost number. There is no loss of
generality in this assumption, since one can always include some
Lagrange multipliers among the canonical variables together with
the constraints that their momenta should vanish.

In subsection \bref{subsec:geom}, we then associate to BRST
quantum mechanics a K\"ahler supermanifold. In particular, the
even symplectic form of ghost number $0$ is determined by the
imaginary part of the non degenerate hermitian inner product. In
appendix {\bf A}, we discuss the geometry of this supermanifold in
terms of complex coordinates. In subsection~\bref{subBFV}, it is
shown that, as for non gauge systems, time evolution in the
supermanifold corresponds to the Hamiltonian flow determined by
the "expectation value" of the BRST invariant Hamiltonian $\mathbf
H=-\half\langle \Psi,\hat H\Psi\rangle$, where $\Psi$ denotes the "string
field". On the supermanifold, the physical state condition then
coincides with the constrained surface determined by the zero
locus of the BRST charge $\mathbf\Omega=-\half\langle \Psi,\hat \Omega
\Psi\rangle$. These constraints are first class, and so is $\mathbf H$.
Furthermore, on the supermanifold, the identification of BRST
closed states up to BRST exact ones corresponds to considering
Dirac observables, i.e., functions defined on the constraint
surface that are annihilated by the Poisson bracket with these
constraints. As has been shown in \cite{Grigoriev:2000zg},
constraints associated to the zero locus of the BRST charge are
special in the sense that the cohomology of the BRST charge itself
provides directly the correct description of these Dirac
observables, without the need to further extend the phase space.
In order to make the paper self-contained, a formal proof adapted
to the BRST charge $\mathbf \Omega$ is provided in appendix {\bf
B}. From the point of view of the symplectic supermanifold, BRST
quantum mechanics becomes thus a classical Hamiltonian BRST system
described by $\mathbf H$ and $\mathbf \Omega$.

Concerning the second step, the proper solution of the master
equation associated to a first order Hamiltonian gauge theory and
its relation to the Hamiltonian BRST formalism is well known
\cite{Batalin88,Siegel:1989nh,Batlle:1989if,Fisch:1989rm,Henneaux:1990ua,%
Batlle:1990zq,Dresse:1990dj,Dresse:1991ba,Grigorian:1991zs}. A
convenient "superfield" reformulation \cite{Grigoriev:1999qz} of
such a master action also exists. These are reviewed in
section~\bref{sec:s2} together with the basic formulas of BRST
operator quantization\footnote{Except for the conventions related
to complex conjugation, we follow closely reference
\cite{Henneaux:1992ig}, to which we refer for further details.}.
In subsection~\bref{subsec:master}, the above results are applied
to derive the master action $\mathbf S$ for the classical
Hamiltonan BFV system of $\mathbf \Omega$ and~$\mathbf H$.

In subsection~\bref{subtencon}, we discuss tensor products of
Hamiltonian BRST quantum mechanical systems at the level of the
associated classical field theories. For later use, the assumption
that the inner product is even is dropped so that the bracket may
be either even or odd. In subsection~\bref{subsec:reinterpret}, it
is shown that the master action $\mathbf S$ associated to $\mathbf
\Omega$ and $\mathbf H$ can be directly obtained from the BRST
charge $\hat \Omega_M$ of the parametrized system: the master
action is given by $\mathbf S=\half\langle \Psi_M,\hat\Omega_M\Psi_M\rangle_M$,
where $\Psi_M$ is the string field of the parametrized system; the
ghost pair of the reparametrization constraint is quantized in the
Schr\"odinger representation so that $\langle \cdot,\cdot\rangle_M$ is odd.

Finally, to complete the last step, we consider in subsection
\bref{ris} the case of systems that are already time
reparametrization invariant and are quantized with an odd inner
product, originating for instance from the Schr\"odinger
representation for the ghosts associated to the mass-shell
constraint (see e.g.~\cite{Siegel:1989ip,Dayi:1993fk}). The master
action $\mathbf S$ is then shown to differ from the original
$\mathbf S_{\rm st}=\half\langle \Psi_{\rm st},\hat\Omega\Psi_{\rm
st}\rangle_{\rm st}$ by two classically trivial pairs, quantized in the
Sch\"odinger representation\footnote{Trivial pairs in string field
theory have been used previously in a different context in
\cite{Siegel:1991zf}.}. More precisely, we show that $\mathbf S$
corresponds to the tensor product of the system described by
$\mathbf S_{\rm st}$ with the system described by the Hamiltonian
BRST charge $\mathbf \Omega_{\rm aux}$ associated to the trivial
pairs. Had these pairs been quantized in the Fock representation
instead, we use the results of subsection~\bref{subtencon} to show
that $\mathbf S$ could have been consistently reduced to $\mathbf
S_{\rm st}$. In the Schr\"odinger representation, however, the
master action $\mathbf S$ involves two more dimensions than
$\mathbf S_{\rm st}$. In subsection~\bref{CS}, we show that
$\mathbf \Omega_{\rm aux}$ is the BRST charge of complex Abelian
Chern-Simons theory. Without additional ingredients, the master
action $\mathbf S$ can then not be directly reduced to $\mathbf
S_{\rm st}$. This is not really surprising since the Fock and the
Schr\"odinger quantization are not unitary equivalent. We conclude
by giving some additional remarks on the BRST charge $\mathbf
\Omega_{\rm aux}$ and the associated master action.

\section{Generalities on BFV and BV formalisms}
\label{sec:s2}

\subsection{Classical Hamiltonian BRST theory}\label{sec2.1}

A Hamiltonian approach to gauge theories involves a symplectic
manifold $\manM_0$ with coordinates $z^A$, constraints $G_{a_0}$,
which we assume for simplicity to be first class and even,
$\pb{G_{a_0}}{G_{b_0}}_{\manM_0}= {C_{a_0b_0}}^{c_0}G_{c_0}$, and
a first class Hamiltonian $H_0$ with
$\pb{H_0}{G_{a_0}}_{\manM_0}={V_{a_0}}^{b_0}G_{b_0}$. The
constraints may be reducible, $Z^{a_0}_{a_1}G_{a_0}=0$, with a
tower of reducibility equations
$Z^{a_{k-1}}_{a_k}Z^{a_{k-2}}_{a_{k-1}}\approx 0$, where $\approx$
means an equality that holds on the constraint surface. Even
though we use a finite-dimensional formulation, this section also
formally applies to field theories by letting the indices $A,a$
range over both a discrete and a continuous set.

In the Hamiltonian BRST approach, the phase space is extended to a
symplectic supermanifold $\manM$ by introducing the ghosts
$\eta^{a_k}$ and the ghost momenta ${\cal P}_{b_k}$ of parity
$k+1$ with $\{{\cal
P}_{a_k},\eta^{b_k}\}_\manM=-\delta_{a_k}^{b_k}$. We take these
variables to be real. Our convention for complex conjugation
involves transposition of variables together with a minus sign
when exchanging two odd variables. On the extended phase space,
the ghost number of a function $A$ that is homogeneous in $\eta^a$
and ${\cal P}_b$ is obtained by taking the extended Poisson
bracket $\pb{\cdot}{\cdot}_\manM$ with the purely imaginary function
\begin{equation}
 {\cal G}=\frac{i}{2}\sum_{k}(k+1)(\eta^{a_k}{\cal
  P}_{a_k}-{\cal P}_{a_k}\eta^{a_k})\,,\qquad
  \pb{A}{\cal G}_\manM=i\gh{A}\, A\,.
\end{equation}
Out of the contraints, one constructs the nilpotent BRST charge of
ghost number 1: \bea \Omega=\eta^{a_0} G_{a_0}+\sum_{k\geq
1}\eta^{a_k}Z_{a_k}^{a_{k-1}}\cP_{a_{k-1}}+\dots\,,
\qquad\half\{\Omega,\Omega\}_\manM=0.\eea Furthermore, the first
class Hamiltonian $H_0$ is extended to the BRST invariant
Hamiltonian $H$ of ghost number $0$ with $\pb{H}{\Omega}_\manM=0$.
Physical quantities such as observables are determined by the BRST
cohomology of the differential $s=\pb{\Omega}{\cdot}_\manM$ in the space
of functions $F(z,\eta,{\cal P})$ on the extended phase space.
Time evolution is generated by the BRST invariant Hamiltonian $H$
according to $\dot F=\pb{F}{H}_\manM$.

\subsection{Master action for first order gauge theories}\label{sec2.2}

In this subsection, we discuss in some details the proper BV
master action for Hamiltonian gauge theories. The reader may wish
to skip these details and go directly to the summary, which is the
only part that is explicitly needed in the rest of the paper.

The information on the symplectic structure, the dynamics and the
constraints of the theory is contained in the extended Hamiltonian
action,
\begin{eqnarray}
S_E[z,\lambda]=\int d\ttime\ (\dot z^A
a^{\manM_0}_A-H_0+\lambda^{a_0} G_{a_0}),\label{1}
\end{eqnarray}
where $\lambda^{a_0}$ are Lagrange multipliers. If the symplectic
two-form is defined by
\begin{equation}
\sigma^{\manM_0}_{AB}=-\ddr{a^{\manM_0}_A}{z^B}-(-1)^{(A+1)(B+1)}\ddr{a^{\manM_0}_B}{z^A},
\end{equation}
the Poisson bracket is determined by
$\pb{z^A}{z^B}_{\manM_0}=\sigma_{\manM_0}^{AB}$ with
$\sigma_{\manM_0}^{AB}\sigma^{\manM_0}_{BC}=\delta^A_C$. Variation
with respect to all the fields $z^A,\lambda^{a_0}$ gives as
equations of motions both the dynamical equations and the
constraints:
\begin{equation}
\dot z^A=\pb{z^A}{H_0}_{\manM_0}\,,\qquad G_{a_0}=0. \label{3}
\end{equation}
A generating set of gauge symmetries for this action is given by
\begin{eqnarray}
\delta_\epsilon z^A&=&\epsilon^{a_0}\{G_{a_0},z^A\}_{{\cal
M}_0},\label{x}
\\
\delta_\epsilon \lambda^{a_0}&=&\dot \epsilon^{a_0}
-\lambda^{c_0}\epsilon^{b_0}
{C_{b_0c_0}}^{a_0}-\epsilon^{b_0}{V_{b_0}}^{a_0}\,,\label{y}
\end{eqnarray}
for some gauge parameters $\epsilon^{a_0}$

In the field-antifield approach, the functional which contains all
the information on the classical action and its gauge algebra is
the proper solution $S$ of the classical master equation,
\begin{eqnarray} \half(S,S)=0.\label{2}
\end{eqnarray}
In the case of the extended Hamiltonian action, the proper
solution $S$ is required to start like the original action
(\ref{1}), to which one couples through the antifields
$z^*_A,\lambda^*_{a_0}$ the gauge transformation (\ref{x}),
(\ref{y}) of the fields with the gauge parameters replaced by the
ghosts $C^{a_0}$. One also needs to couple the terms containing
the Lagrangian reducibility operators, (which are determined by
the Hamiltonian reducibility operators $Z^{a_{k-1}}_{a_k}$) and
introduce associated ghosts for ghosts and their antifields. The
antifields can be chosen to be real and are defined to be
canonically conjugate to the fields with respect to the
antibracket $(\cdot,\cdot)$. Additional terms in $S$ are then
uniquely determined by the master equation (\ref{2}), up to
anticanonical transformations in the antibracket. The proper
solution $S$ associated to \eqref{1} can then be shown to be given
by
\begin{eqnarray}
S[z,z^*,\lambda,\lambda^*,\eta,\eta^*]=\int d\ttime\ \Big(\dot z^A
a^{\manM_0}_A
+\sum_{k\geq 0}\dot\eta^{a_k}{\cal P}_{a_k} -H\nonumber\\
-z^*_A\pb{z^A}{\Omega}_\manM-\sum_{k\geq 0}[\lambda^{a_k}\pb{{\cal
P}_{a_k}}{\Omega}_\manM+
\eta^*_{a_k}\pb{\eta^{a_k}}{\Omega}_\manM]\Big),
\label{eq:maction2}
\end{eqnarray}
where the identifications $C^{a_k}=\eta^{a_k},
C^*_{a_k}=\eta^*_{a_k}$ and ${\cal P}_{a_k}=-\lambda^*_{a_k}$ have
been made\footnote{For later convenience, some signs have been
changed in equations \eqref{1}, \eqref{x} and \eqref{y} with
respect to those of \cite{Henneaux:1992ig}. In
\eqref{eq:maction2}, they imply the change
$\Omega\rightarrow-\Omega$.}

Usually, in order to fix the gauge, one introduces a nonminimal
sector, containing antighosts in ghost number $-1$, their momenta
in ghost number $1$ and auxiliary fields in ghost number $0$.
Then, a gauge fixing fermion $\Upsilon$ in ghost number $-1$ that
depends only on the fields is chosen. The choice of $\Upsilon$ is
determined by the requirement that there be no more gauge
invariance in the dynamics generated by the nonminimal master
action obtained after application of the anticanonical
transformation generated by $\Upsilon$ and after setting to zero
the transformed antifields. This gauge fixed action can be taken
as a starting point for a path integral quantization and the
partition function can (formally) be shown to be independent of
the choice of $\Upsilon$.

For the master action \eqref{eq:maction2}, it is possible to fix
the gauge without introducing a nonminimal sector: indeed, by
considering the anticanonical transformation which consists in the
exchange of fields and antifields for the sector of the Lagrange
multipliers, \bea (\lambda^{a_k},\lambda^*_{a_k})\longrightarrow
(-\lambda^*_{a_k},\lambda^{a_k})\equiv ({\cal P}_{a_k},{\cal
P}^{*a_k}), \eea the equations of motion are in first order form.
The new fields are then the fields $z^\alpha=(z^A,\eta^{a_k},{\cal
  P}_{a_k})$ that are naturally associated with the Hamiltonian BRST
formalism. The antibracket for two functionals $A[z,z^*],B[z,z^*]$
is defined by \bea (A,B)[z,z^*]=\int d\ttime\Big [
\vddr{A}{z^\alpha(\ttime)}\vddll{B}{z^*_\alpha(\ttime)}-
(z^\alpha\longleftrightarrow
z^*_\alpha)\Big].\label{antibracket}\eea The solution $S$ of the
master equation \eqref{3} can be rewritten in a compact way as
\begin{equation}
\label{eq:maction} S[z,z^*]=\int d\ttime\ (\dot z^\alpha
a^\manM_\alpha -H-\pb{z^*_\alpha z^\alpha}{\Omega}_\manM).
\end{equation}
An additional gauge fixing generated by the fermion $\Upsilon=\int
d\ttime\ K(z)$ can then be considered. Its effect is to change the
BRST invariant Hamiltonian by a BRST exact term, $H\rightarrow
H+\{K,\Omega\}$. Note that after putting to zero the antifields
$z^*_\alpha$, the constraint equations \eqref{3} are no longer
imposed as equations of motions since the associated fields have
been put to zero.

In the following, we will not put to zero the antifields obtained
after a canonical transformation generated by $\Upsilon$. This is
because during the renormalization process, the antifields allow
to conveniently control the Ward identities due to BRST invariance
under the form of the Zinn-Justin equation for the effective
action. To lowest order in $\hbar$, it is the antifield dependent
BRST cohomology of the differential $s=(S,\cdot)$ that controls
gauge invariance on the quantum level. This cohomology is
invariant under canonical transformations and the introduction of
a non minimal sector. Hence, from this point of view, one can
forget about gauge fixing and directly discuss the cohomology
associated to the master action \eqref{eq:maction}. In turn, this
cohomology computed in the space of functions in the fields
$z^\alpha$, the antifields $z^*_\alpha$ and their (space)time
derivatives can be shown to be isomorphic to the Hamiltonian BRST
cohomology of the differential $s_\Omega=\{\Omega,\cdot\}_\cM$ in
the space of functions in $z^\alpha$ (and their spatial
derivatives). In the space of local functionals, which is the
relevant space in the context of renormalization, the relation
between with the Hamiltonian BRST cohomology is more involved
\cite{Barnich:1996mr}.

\bigskip

\noindent
{\bf Summary:}

\noindent From the above construction of the solution
\eqref{eq:maction} to the classical master equation, we can learn
the following. Suppose that the following data is given:
\begin{itemize}
\item a (super) phase space with coordinates $z^\alpha$ and
  symplectic $2$ form generated by $a_\alpha(z)$ with associated
  Poisson bracket
  $\{z^\alpha,z^\beta\}=\sigma^{\alpha\beta}_{\manM}(z)$,

\item a ghost number grading ${\cal G}$ on the phase space,

\item a nilpotent BRST charge $\Omega$ in ghost number $1$,
whose cohomology determines the
  physically relevant quantities on the phase space,

\item a BRST invariant Hamiltonian $H$ in ghost number $0$
determining the time evolution.
\end{itemize}
Then, in the space of functionals in the fields $z^\alpha(\ttime)$
and additional independent antifields $z^*_\alpha(\ttime)$ of
ghost number $-gh(z^\alpha)-1$ equipped with the antibracket given
by \eqref{antibracket}, the proper solution of the master equation
is given by \eqref{eq:maction}. In order to recover the gauge
invariant equations of motion (including the constraints) after
putting to zero the antifields, the interpretation of which are
the fields and which are the antifields should be reversed for the
fields in negative ghost number, $({\cal P}_a,{\cal P}^{*a})\equiv
(-\lambda^*_a,\lambda^a)\longrightarrow (\lambda^a,\lambda^*_a)$.

\subsection{Superfield reformulation}\label{sec2.3}

A superfield reformulation \cite{Grigoriev:1999qz} of the master
action \eqref{eq:maction} is achieved by introducing an additional
Grassmann odd variable $\theta$ of ghost number one.

Given an extended  phase space $\manM$, one associates a space
$\Sigma$ of maps $z^\alpha=z_S^\alpha(\ttime,\theta)$ from the
$(1|1)$-dimensional superspace spanned by $\ttime$ and $\theta$ to
$\manM$.  This space is a super-extension of the space of field
histories $z^\alpha(\ttime)$ (maps from $\ttime$ to $\manM$).
Functionals on $\Sigma$ can be identified with functionals in the
fields and antifields of the previous section. Indeed, one can
expand $z^\alpha_S(\ttime,\theta)$ into components
\begin{equation}
  z_S^\alpha(\ttime,\theta)=z^\alpha(\ttime)+\theta{z^*}_\beta(\ttime)
  \sigma^{\beta\alpha}_\manM(z(\ttime)) \,,\label{comp}
\end{equation}
which is consistent with the various ghost number assignments. To
every functional ${\cal A}[z_S]$ one can associate the functional
$A[z,z^*]$ obtained by using this expansion. Conversely, to every
functional $A[z,z^*]$ corresponds the functional ${\cal
A}[z_S]=A[\int d\theta\theta z_S,\int d\theta z_S\sigma(z_S)]$.

Functionals on $\Sigma$ are equipped with the odd Poisson bracket
\bea ({\cal A},{\cal B})[z_S]=(-1)^{\p{\cA}+1}\int d\ttime d\theta
\vddr{\cA}{z_S^\alpha(\ttime,\theta)}
\sigma^{\alpha\beta}_\manM(z_S(\ttime,\theta))
\vddll{\cB}{z_S^\beta(\ttime,\theta)}, \eea with $({\cal A},{\cal
B})[z+\theta z^*\sigma^{-1}]=(A,B)[z,z^*]$. This Poisson bracket
is odd and of ghost number~$1$. Functional derivatives are defined
as \bea \delta\cA=\int dtd\theta\, \delta
z^\alpha(t,\theta)\vddll{\cA}{z^\alpha(t,\theta)}= \int
\vddr{\cA}{z^\alpha(t,\theta)}\delta
z^\alpha(t,\theta)\,dtd\theta.\eea
The master action ${\cal S}[z_S]$ corresponding to $S[z,z^*]$
given in \eqref{eq:maction} can then be written as
\begin{equation}
 {\cal S}[z_S]=\int\, d\ttime d\theta\,
 \Big[ Dz_S^\alpha a^\manM_\alpha(z_S)-\theta H(z_S)-\Omega(z_S)\Big]\,,
\end{equation}
with $D=\theta\dd{}{\ttime}$.

The superfield reformulation regroups fields and antifields in
convenient supermultiplets so that the antibracket is induced by
the extended Poisson bracket.

\subsection{BRST operator quantization}\label{BRSTop}

The BRST operator quantization consists in realizing the functions
on the extended phase space as linear operators in a super Hilbert
space ${\cal
  H}$ together with the correspondence rule $[\hat A,\hat
B]=i\hbar\widehat{\{A,B\}} +O(\hbar^2)$, where $[\cdot,\cdot]$
denotes the graded commutator and $A,B$ are phase space functions
with associated linear operators $\hat A,\hat B$.

These rules imply in particular that $\half[\hat \Omega,\hat
\Omega]=O(\hbar^2)=[\hat H,\hat \Omega]$. In the following, we
assume that we are in the non anomalous case, where
\begin{eqnarray}
\half[\hat\Omega,\hat\Omega]=0,\qquad
[\hat{H},\hat\Omega]=0,\label{2.16}
\end{eqnarray}
and $\hat\Omega$, $\hat H$ are hermitian operators in the inner
product $\inner{\psi}{\phi}$, which is non degenerate but not
necessarily positive definite and makes the real classical
variables hermitian operators. Furthermore, we take $\hbar=1$. For
a super Hilbert space, \bea
\overline{\inner{\psi}{\phi}}&=&(-1)^{\p{\psi}{\p{\phi}}}\inner{\phi}{\psi},\\
\inner{\psi}{\hat A\phi} &=&(-1)^{\p{A}{\p{\psi}}}\inner{\hat
A^\dagger\psi}{\phi},\eea where $\p{\phi},\p{A}$ denotes the
Grassmann parity of the state, respectively the operator $\hat A$.
The relation to standard Hilbert space with even and odd elements,
for which the above formulas do not involve sign factors, is
explained for instance in \cite{DeligneFreed1}.

In what follows, we are not interested in a probabilistic
interpretation of the quantum theory, but rather in an associated
classical field theory. This is the reason why we are not
concerned here with questions related to the normalizability of
states or to the infinite dimensionality of the Hilbert space.

The ghost number of an operator is obtained by taking the graded
commutator (from the left) with the antihermitian operator
$\hat{{\cal G}}$. We assume that ${\cal H}$ splits as a sum of
eigenstates of $\hat{{\cal G}}$,  ${\cal H}=\oplus_p {\cal H}_p$
with $\hat{{\cal G}}\psi_p=p\psi_p$ for $\psi_p\in {\cal H}_p$. It
then follows from the antihermiticity of $\hat{{\cal G}}$ that
$\inner{\psi_p}{\phi_{p^\prime}}\neq 0$ only if $p+p^\prime=0$.
This means that the ghost number of the scalar product
$\inner{}{}$ is zero. The ghost number $p$ of a state can be shown
to be $p=p_0+k$ for some integer $k$ with $p_0=0$ or $p_0=\half$.

The case where $p_0=\half$ arises if the number of independent
constraints is odd. In this case, one can include some of the
Lagrange multipliers $\lambda^a$ and their momenta $b_a$ among the
canonical variables, $\{\lambda^a,b_b\}=\delta^a_b$, together with
the new constraint $b_a\approx 0$. On the level of the classical
BRST formalism, this implies adding to the extended phase space
the antighosts $\bar C_a$ of ghost number $-1$ and their momenta
$\rho^a$ of ghost number $1$, $\{\rho^a,\bar C_b\}=-\delta^a_b$.
All these variables are chosen to be real. The BRST charge of the
system is then modified by the addition of the non minimal piece
$\Omega^{{\rm nm}}= \rho^ab_a$. Hence, by adding the
cohomologically trivial pairs $(\lambda^a,b_a)$, $(\rho^a,\bar
C_a)$, one can always assume $p_0=0$, which is what we do unless
otherwise specified. We also assume that the inner product is
even, $\inner{\psi}{\phi}=0$ if $\psi$ and $\phi$ are of opposite
parity.

Physical operators are described by hermitian operators $\hat A$
such that
\begin{eqnarray}
[\hat A,\hat \Omega]=0,
\end{eqnarray}
where two such operators have to be identified if they differ by a
BRST exact operator
\begin{eqnarray}
\hat A\sim \hat A+[\hat B,\hat \Omega].
\end{eqnarray}
These two equations define the BRST operator cohomology.

Similarily, physical states are selected by the condition
\begin{eqnarray}
\hat \Omega\psi=0.\label{physstate}
\end{eqnarray}
Furthermore, BRST exact states should be considered as zero, or
equivalently, states that differ by a BRST exact ones should be
identified
\begin{eqnarray}
\psi\sim\psi+\hat \Omega\chi.\label{idofphys}
\end{eqnarray}
These two equations define the BRST state cohomology.

Finally, time evolution is governed by the Schr\"odinger equation
\begin{eqnarray}
i \frac{d \psi}{d \ttime}= \hat H \psi\label{4}.
\end{eqnarray}

\section{BFV and BV formalisms for BRST first quantized gauge
  systems}
\label{sec:BFV-BV}

\subsection{Geometry of BRST quantum mechanics}\label{subsec:geom}

Let $\{e_a\}$ be a basis over ${\mathbb R}$ of the graded Hilbert
space ${\cal H}$ such that the basis vectors are of definite
Grassmann parity $\p{a}$ and ghost number $\gh{e_a}$. A general
vector can be written as $\psi=e_ak^a$, with $k^a\in {\mathbb R}$.
To each $e_a$, one associates a real variable $\Psi^a$ of parity
$\p{a}$ and ghost number $-\gh{e_a}$. These variables are
coordinates of a supermanifold $\cM_\cH$ associated to ${\cal H}$.
The algebra of real valued functions on this supermanifold is
denoted by $\mathfrak G$. Introducing the right module
$\cH_{\mathfrak G}={\cal H}\otimes \mathfrak G$
\cite{Gaberdiel:1997ia}, the "string field" appears as the
particular element $\Psi=e_a\Psi^a$ of this module. At this stage,
it is even and of total ghost number $0$.

The sesquilinear form $\inner{\cdot}{\cdot}$ on $\cH$ can be extended to
elements $\psi f$ and $\phi g$ of $\cH_{\mathfrak G}$, with
$f(\Psi),g(\Psi)\in \mathfrak G$ by the rule \bea \inner{\psi
f}{\phi g}=(-1)^{\p{f}\p{\phi}}\inner{\psi}{\phi}fg\,.\eea A
linear operator $\hat A$ on $\cH$ is naturally extended to
$\cH_{\mathfrak G}$: $\hat A(\psi f)=(\hat A\psi)f$.

The real and imaginary parts of this inner product, \bea
\inner{\psi}{\phi}=g(\psi,\phi)+i\omega(\psi,\phi) \eea are
respectively graded symmetric and graded skew symmetric, \bea
g(\psi,\phi)&=&(-1)^{\p{\psi}\p{\phi}}g(\phi,\psi)\,,\\
\omega(\psi,\phi)&=&-(-1)^{\p{\psi}\p{\phi}}\omega(\phi,\psi).
 \eea
The forms $g(\cdot,\cdot)$ and $\omega(\cdot,\cdot)$ are extended
to $\cH_{\mathfrak G}$ in the same way as $\inner{\cdot}{\cdot}$.

If $\cH$ is considered as a superspace over real numbers, both
$g(\psi,\phi)$ and $\omega(\psi,\phi)$ are ${\mathbb R}$-bilinear
forms on $\cH$. The complex structure $\hat J$ is the linear
operator that represents multiplication by $i$. As a consequence,
\begin{gather}
g(\hat J\phi, \hat J \psi)= g(\phi, \psi)\,, \qquad  \omega(\hat
J\phi, \hat J \psi)=
\omega(\phi, \psi)\,.\\
g(\hat J \phi, \psi) = \omega(\phi,\psi)\,.
\end{gather}
Furthermore, the operator $\hat J$ commutes with $\mathbb
C$-linear operators.

Introducing the coefficients
$\omega_{ab}=(-1)^{\p{a}}\omega(e_a,e_b)$ and defining
$\omega^{ab}$ through $\omega^{ab}\omega_{bc}=\delta^a_c$, an even
graded Poisson bracket on $\mathfrak G$ of ghost number $0$ is
defined by
\begin{equation}
\pb{f}{g}= \ddr{f}{\psi^a}\omega^{ab}\ddl{\psi^b}{g}\,.
\end{equation}

To each antihermitian operator $\hat A$, one associates a real
quadratic function $F_{\hat A}(\Psi)\in {\mathfrak G}$ by
\begin{equation}\label{eq:iso}
  F_{\hat A}(\Psi)=\half\inner{\Psi}{-\hat J\hat A\Psi}\,.
\end{equation}
Antihermiticity implies that $F_{\hat
A}(\Psi)=\half\omega(\Psi,\hat A\Psi)=-\half\omega(\hat
A\Psi,\Psi)$. This map is an homomorphism from the super Lie
algebra of antihermitian operators to the super Lie algebra of
quadratic real functions in $\mathfrak G$ equipped with the
Poisson bracket
\begin{equation}
\pb{F_{\hat A}}{F_{\hat B}}=F_{\commut{\hat A}{\hat B}}\,.
\end{equation}
The map is compatible with parity and ghost number assignments,
$\gh{F_{\hat A}}=\gh{\hat A}$, $\p{F_{\hat A}}=\p{\hat A}$.

For hermitian operators, we define $\mathbf A(\Psi)=F_{-\hat J\hat
A}$. Because of hermiticity \bea \mathbf
A(\Psi)=-\half\inner{\Psi}{\hat A\Psi}=-\half g(\Psi,\hat
A\Psi)=-\half g(\hat A\Psi,\Psi).\eea Furthermore, the properties
of $\hat J$ imply that \bea \{{\mathbf A},{\mathbf
B}\}=-\half\inner{\Psi}{[\hat A,\hat B]\Psi}. \eea In particular,
for the hermitian BRST charge $\hat \Omega$ and the hermitian BRST
invariant Hamiltonian $\hat H$, equations \eqref{2.16} imply
\begin{equation}
  \begin{gathered}
        \half\{{\mathbf \Omega},{\mathbf \Omega}\}=0,\qquad
\{{\mathbf H},{\mathbf \Omega}\}=0,\label{fund}
\end{gathered}
\end{equation}
where ${\mathbf H},{\mathbf \Omega}$ are of total ghost numbers
$0$ and $1$ respectively.

\subsection{BRST quantum mechanics as classical BFV system}\label{subBFV}

The Schr\"odinger equation in terms of $\Psi^a$ can be written as
\bea \frac{d\Psi^a}{d\ttime}=-(\hat J\hat
H\Psi)^a=\{\Psi^a,\mathbf H \},\eea so that time evolution of
elements $f(\Psi)\in \mathfrak G$ is determined by the Hamiltonian
flow of $\mathbf H$, \bea \frac{df}{d\ttime}=\{f,\mathbf H\}.\eea

On $\cM_\cH$, the physical state condition \eqref{physstate}
defines a submanifold, the constraint surface determined by  \bea
\hat \Omega^a_b\Psi^b\approx 0.\eea Because \bea \{\cdot,\mathbf
\Omega\}=\ddr{\cdot}{\Psi^a}(-\hat J\hat\Omega\Psi)^a\,,\eea the
constraint surface can be identified with the zero locus of the
Hamiltonian vector field associated to $\mathbf \Omega$, \bea
\mathbf G^a\equiv \{\Psi^a,\mathbf \Omega\}\approx 0\,.  \eea By
using the graded Jacobi identity for the Poisson bracket and
taking \eqref{fund} into account, these constraints are easily
shown to be first class, \bea \{\mathbf G^a,\mathbf G^b\}=
\{\{\Psi^a,\mathbf \Omega\},\{\Psi^b,\mathbf
\Omega\}\}=\{\{\{\Psi^a,\mathbf \Omega\},\Psi^b\},\mathbf
\Omega\}\nonumber\\=\ddr{\{\{\Psi^a,\mathbf
\Omega\},\Psi^b\}}{\Psi^c}\mathbf G^c\approx 0 .\eea Furthermore,
since $\mathbf \Omega$ is quadratic in $\Psi$, these constraints
are in fact abelian, \bea \{\mathbf G^a,\mathbf G^b\}=0. \eea

On $\cM_\cH$, the identification~\eqref{idofphys} of
states up to BRST exact ones, corresponds to taking functions on
$\mathfrak G$ that are annihilated by the distribution generated
by $\hat\Omega^a_b\ddl{\Psi^a}{\cdot}$. This distribution is
equivalently generated by the adjoint action of the constraints
$\{\mathbf G^a,\cdot\}$. The Hamiltonian $\mathbf H$ is also first
class, \bea \{\mathbf H,\mathbf G^a\}=\{\{\mathbf
H,\Psi^a\},\mathbf \Omega\}= \ddr{\{\mathbf
H,\Psi^a\}}{\Psi^b}\mathbf G^b.\eea

Hence, from the point of view of $\cM_\cH$, BRST quantum mechanics
becomes a classical constraint Hamiltonian system. According to
the Dirac theory, an observable is a function $f(\Psi)\in
\mathfrak G$ such that $\{f,\mathbf G^a\}\approx 0$. Two such
functions should be considered equivalent if they coincide on the
constraint surface, $f\sim f+\lambda_a \mathbf G^a$. Equivalence
classes of observables then form a Poisson algebra with respect to
the induced bracket.

The classical Hamiltonian BRST approach described in subsection
\ref{sec2.1} consists in extending the phase space in order to
encode this Poisson algebra in terms of the cohomology of a BRST
charge. This will however not be straightforward in the case of
the zero locus constraints $\mathbf G^a$, because they are
reducible due to the nilpotency of $\Omega$, and for the obvious
reducibility operators, they are infinitely reducible.

In fact, it turns out that for the zero locus constraints $\mathbf
G^a$, there is actually no need to extend the phase space. Indeed,
the Poisson algebra of equivalence classes of observables is
isomorphic to the cohomology of the BRST charge $\mathbf \Omega$
itself, equipped with the induced Poisson bracket. This has been
shown in \cite{Grigoriev:2000zg}, where constraint systems
originating from the zero locus of a generic Hamiltonian BRST
differential have been analyzed. A proof adapted to the particular
BRST charge $\mathbf \Omega$ is given in appendix {\bf B}.

As a side remark, let us note that treating the zero locus of the
BRST charge as a constraint surface is analogous to considering
the master action $S$ as a classical action; in this case, the
zero locus of the BRST differential $s=(S,\cdot)$ is the
stationary surface associated to $S$ (see e.g.~
\cite{Sen:1994ic,Grigoriev:1998gn,Grigoriev:2000zg}).

\subsection{Proper master action for BRST quantum mechanics}\label{subsec:master}

According to subsection \ref{sec2.2}, the solution of the master
equation associated to the classical Hamiltonian BRST system on
the phase space $\cM_\cH$ described by $\mathbf H$ and $\mathbf
\Omega$ is given by \bea {\mathbf S}[\Psi,\Psi^*]=\int d\ttime\
[\half\omega(\Psi,\frac{d}{d\ttime}\Psi)-\mathbf
H-\{\Psi^*_a\Psi^a,\mathbf \Omega\}]\, \eea which can be written
as
\begin{multline} \label{prop} {\mathbf S}[\Psi,\Psi^*]=
\half \int d\ttime \Big  (-i \langle \Psi,\frac{d}{d\ttime}
\Psi\rangle+\langle \Psi,\hat H\Psi\rangle -\cr-\langle \tilde \Psi^*,\hat \Omega \Psi
\rangle+\langle \Psi,\hat \Omega \tilde \Psi^* \rangle\Big )\,, \end{multline} where
$\tilde\Psi^{*a}=\Psi^*_b\omega^{ba}$ and $\tilde\Psi^*=
e_a\tilde\Psi^{*a}$. As explained in section \ref{sec2.2}, the
role of fields and antifields has been exchanged for those fields
that are in strictly negative ghost numbers.

According to subsection \ref{sec2.3}, we now introduce \bea
\Psi_S^a(\ttime,\theta)=\Psi^a(\ttime)+\theta \tilde
\Psi^{*a}(\ttime), \eea and also the ghost number $0$ object \bea
\Psi_S=e_a \Psi_S^a(\ttime,\theta).\label{ssf}\eea The proper
solution \eqref{prop} can then be written as
\begin{equation} {\mathbf S}[\Psi_S]=
\half \int d\ttime d\theta\, \Big  (-i \theta \langle \Psi_S,
\frac{d}{d\ttime} \Psi_S\rangle+\,\theta \langle \Psi_S,\hat H\Psi_S\rangle
+\langle \Psi_S,\hat \Omega \Psi_S
\rangle\Big )
\,. \end{equation} By construction, it satisfies the master
equation with respect to the antibracket \eqref{comp}, with
$z^A_S(\ttime,\theta)$ replaced by $\Psi^a_S(\ttime,\theta)$,
\bea
({\cal A},{\cal B})[\Psi_S]=(-1)^{\p{A}+1}\int d\ttime d\theta
\vddr{\cal A}{\Psi_S^a(\ttime,\theta)} \omega^{ab}
\vddll{\cal B}{\Psi_S^b(\ttime,\theta)}.\label{abr}
\eea
For later purposes, it will be useful to rewrite the master action as
\begin{equation} {\mathbf S}[\Psi_S]=
\half \int d\ttime d\theta\, \langle \Psi_S,(-i \theta
\frac{d}{d\ttime}+\theta \hat H+\hat \Omega)\Psi_S\rangle \,.\label{ma1}
\end{equation}

\section{Master action and time reparametrization invariance}\label{ri}

\subsection{Tensor constructions}\label{subtencon}

Given two first quantized BRST systems with super-Hilbert spaces
$\cH_i,\,i=1,2$, their respective BRST charges $\hat\Omega_i$ and
Hamiltonians $\hat H_i$, the tensor product
$\cH_T=\cH_1\otimes_{\mathbb C}\cH_2$ is again a super Hilber
space. For later use, we do not assume in this section that the
inner products on $\cH_i$ are even, but we allow them to be of
arbitrary parity $\varepsilon_i$. We also admit the possibility of
fractionalization of the ghost number. The Grassmann parity and
the ghost number of the state $\phi_1\densor\phi_2$ is naturally
$\p{\phi_1}+\p{\phi_2}$ and $\gh{\phi_1}+\gh{\phi_2}$
respectively. The inner product on $\cH_T$ is determined by
\begin{equation}
  \inner{\phi_1\densor\phi_2}{\psi_1\densor\psi_2}_{T}=
(-1)^{\p{\psi_1}\p{\phi_2}} \inner{\phi_1}{\psi_1}_{1}
\inner{\phi_2}{\psi_2}_{2}\,.
\end{equation}
It is of parity $\varepsilon_1+\varepsilon_2$ and non degenerate
(for $\cH_T$ considered as a complex space) provided the ones on
$\cH_1$ and $\cH_2$ are.

Linear operators $\hat A_i$ on $\cH_i$ determine a linear operator
$\hat A_T$ on $\cH_T$ by
\begin{equation}
\hat A_T(\phi_1\densor\psi_2)=(\hat A_1\densor {\mathbf
1}+{\mathbf 1}\densor \hat A_2)(\phi_1\densor\psi_2)= (\hat
A_1\phi_1)\densor\psi_2+(-1)^{\p{A_2}\p{\phi_1}}\phi_1\densor(\hat
A_2 \psi_2)\,.
\end{equation}
The various definitions imply that
\begin{equation}
\hat A_T^\dagger=(\hat A_1\densor {\mathbf 1}+ {\mathbf 1}\densor
\hat A_2)^\dagger= \hat A^\dagger_1\densor {\mathbf 1}+ {\mathbf
1}\densor \hat A^\dagger_2\,,
\end{equation}
and
\begin{equation}
\commut{\hat A_T}{\hat B_T}= \commut{\hat A_1}{\hat B_1}\densor
{\mathbf 1}+{\mathbf 1}\densor \commut{\hat A_2}{\hat B_2} \,.
\end{equation}
In particular, the BRST charges $\hat \Omega_i$ and the BRST
invariant Hamiltonians $\hat H_i$ determine hermitian operators
$\Omega_T$ and $\hat H_T$ such that $\frac{1}{2}[\hat
\Omega_T,\hat \Omega_T]=0$ and $[\hat H_T,\hat \Omega_T]=0$.
Furthermore,
\begin{equation}
H(\hat\Omega_T,\cH_T)=H(\hat\Omega_1,\cH_1){\densor}_{\mathbb C}
H(\hat\Omega_2,\cH_2).
\end{equation}
The formal proof is elementary and given in appendix {\bf C}.

If $\{e_\alpha,e_{\bar\alpha}\}$ is a basis of $\cH_1^{\mathbb C}$
(see appendix {\bf A}), while $\{E_\Lambda,E_{\bar\Lambda}\}$ is a
basis of $\cH^{\mathbb C}_2$, then $\{e_\alpha \densor
E_\Lambda,e_{\bar\alpha} \densor E_{\bar\Lambda}\}$ is a basis of
$\cH^{\mathbb C}_T$. For these basis vectors, one can consider the
complex coordinates $\Psi^{\alpha\Lambda}$ and $\bar\Psi^{\9\alpha
\9\Lambda}$ for the supermanifold $\manM_T$ associated to the
superspace $\cH_T$ and also the associated complex valued
functions $\mathfrak G_T^{\mathbb C}$. The string fields can now
be defined by
\bea {\Psi_T}=(e_\alpha\otimes
e_\Lambda)\Psi^{\alpha\Lambda}+(e_{\bar\alpha}\otimes
E_{\bar\Lambda})\Psi^{\bar\alpha\bar\Lambda}\,.
\eea
The functions
\bea\mathbf
\Omega_T=-\half\langle \Psi_T,\hat\Omega_T\Psi_T\rangle_T,\qquad\mathbf
H_T=-\half\langle \Psi_T,\hat H_T\Psi_T\rangle_T\,,
\eea
satisfy
\bea
\half
\{\mathbf \Omega_T,\mathbf \Omega_T\}_T=0=\{\mathbf H_T,\mathbf \Omega_T\}_T\,,
\eea
where $\pb{\cdot}{\cdot}_T$ denotes a Poisson bracket or antibracket
on $\mathfrak G_T^{\mathbb C}$ determined by imaginary part of $\langle \cdot,\cdot\rangle_T$.
Note, however, that when $\langle \cdot,\cdot\rangle_T$ is odd, so is
the Poisson bracket $\{\cdot,\cdot\}_{\mathfrak G^T_{\mathbb C}}$.
In this case, it is also called "antibracket" and ${\mathbf \Omega_T}$ is a master action.

If $\{e_\theta,e_{\bar\theta}\}$, $\{E_\Theta,E_{\bar\Theta}\}$
are bases over $\mathbb C$ of $H(\hat \Omega_1,\cH^{\mathbb
C}_1)$, $H(\hat \Omega_2,\cH_2^{\mathbb C})$, it follows from
appendix {\bf B} and appendix {\bf C} that the cohomology of
$\{\cdot,\mathbf \Omega_T\}_{\mathfrak G_T}$ is isomorphic to real
functions in $\Psi^{\theta\Theta}$ and $\Psi^{\bar\theta
\bar\Theta}$. In particular, if $H(\hat\Omega_2,\cH_2)$ is a one
dimensional space over $\mathbb C$ with basis vector $E$ such that
$\gh{E}=0,\p{E}=0$, then $H(\Omega_T,\cH_T)\simeq
H(\Omega_1,\cH_1)$. It also follows that the cohomology of
$\{\cdot,\mathbf \Omega_T\}_{\mathfrak G_T}$ in ${\mathfrak G_T}$
is isomorphic to that of $\{\cdot,\mathbf \Omega_1\}_{\mathfrak
G_1}$ in ${\mathfrak G_1}$.

Suppose that $\cH_2$ contains a quartet or a null doublet. Let
$\Lambda=(i,\Lambda^\prime)$, where $i$ runs over the states of
the quartet or the null doublet. Not only do these states not
contribute to the cohomology, but they can also be consistently
eliminated from $\mathbf \Omega_T$ by reducing the string field
used in the construction of $\mathbf \Omega_T$ to \bea
\Psi^\prime_T=(e_\alpha\otimes
E_{\Lambda^\prime})\Psi^{\alpha{\Lambda^\prime}}+(e_{\bar\alpha}\otimes
E_{\bar\Lambda^\prime})\Psi^{\bar\alpha\bar\Lambda^\prime}\,.\eea
This elimination is algebraic. If the parity of $\langle \cdot,\cdot\rangle_T$
is odd, it corresponds to the elimination of "generalized
auxiliary fields" of the master action discussed in
\cite{Dresse:1990dj}. In the case where the parity
$\langle \cdot,\cdot\rangle_T$ is even, it is an Hamiltonian analogue of this
concept.

\subsection{Reinterpretation of master action for BRST quantum mechanics}\label{subsec:reinterpret}

Consider now the super Hilbert space $\cH_{t,\theta}$ obtained by
quantizing the phase space $(t,p_0),(\theta,\pi)$ in the
Schr\"odinger representation. The wave functions are
$\varphi(t,\theta)=\varphi_0(t)+\theta\varphi_1(t)$, while the
inner product is given by
\begin{equation}
\langle \varphi,\varrho\rangle_{t,\theta}=\int dt d\theta\, \bar
\varphi(t,\theta)\varrho(t,\theta)\,.
\end{equation}
The ghost number operator is given by $\hat {\cal
G}_{t,\theta}=\theta\dl{\theta}-\half$ so that
$\gh{\varphi_0(t)}=-\half$, $\gh{\theta\varphi_1(t)}=\half$ and we
take $\p{\phi_0(t)}=0$, $\p{\theta\varphi_1(t)}=1$ so that the
inner product is Grassmann odd and of ghost number $0$. For the
super Hilbert space $\cH_M=\cH\otimes_{\mathbb C}\cH_{t,\theta}$,
where $\cH$ is an even super Hilbert space, states are of the form
$e_ak_0^a(t)+ e_a\theta k_1^a(t)$. Real coordinates on the
supermanifold $\cM_{\cH_M}$ associated to $\cH_M$ can be chosen as
$k_0^a(t)\rightarrow \Psi^a(t)$, $k_1^a(t)\rightarrow
\tilde\Psi^{*a}(t)$. The ghost number $0$ object $\Psi_S$
introduced in \eqref{ssf} can now be identified with the string
field $\Psi_M$ associated to $\cH_M$, $\Psi_S\equiv\Psi_M$. The
odd inner product on $\cH_M$ is denoted by $\langle \cdot,\cdot\rangle_M$ and
extended to $\cH_M\otimes\mathfrak G_M$, where $\mathfrak G_M$ is
the algebra of real functions in $\Psi^a(t),\tilde\Psi^{*a}(t)$.
The master action \eqref{ma1} can now be written as \bea {\mathbf
S}[\Psi_M]&=&\half\langle \Psi_M,\hat
\Omega_M\Psi_M\rangle_M\label{ma2}\,,\\\hat \Omega_M&=&\hat\theta(\hat
p_0+\hat H)+\hat \Omega. \eea In this case, the imaginary part of
this inner product determines an odd symplectic structure on
$\cM_{\cH_M}$. Its inverse is an antibracket, which coincides with
\eqref{abr}, up to ghost number assignments discussed below. Only
when the BRST invariant Hamiltonian $\hat H$ vanishes is $\hat
\Omega_M$ the tensor product of the BRST charges $\hat \Omega$ and
$\hat\theta\hat p_0$.

From the expression of $\hat\Omega_M$, we can deduce how to arrive
directly at \eqref{ma2}: make the original classical Hamiltonian
system time reparametrization invariant by including the time
$\ttime$ and its conjugate momentum $p_0$ among the canonical
variables and adding the first class constraint $p_0+H_0\approx
0$. The BRST charge for this system is then given by $\Omega_M$,
with $(\theta,\pi)$ the "time reparametrization" ghost pair
associated with this new constraint. The quantization of this
system then leads directly to $\cH_M$ with its odd inner product
and the master action \eqref{ma2}.

From the point of view of $\cH_M$, the ghost numbers of fields and
antifields are half integer and differ from the standard ones by
one half. In particular physical fields are those at ghost number
$\half$). This originates from the additional ghost number
operator $\hat {\cal G}_{t,\theta}$ and our convention for the
string field ghost number. Indeed, the fields $\Psi^a(t), \tilde
\Psi^{*a}$ now have ghost numbers $\half-\gh{e_a}$ and
$-\half-\gh{e_a}$ instead of $-\gh{e_a}$ and $-1-\gh{e_a}$, which
is the natural assignment from the point of view of the BFV system
associated to BRST quantum mechanics, and, as explained in
subsection \bref{subsec:master}, also leads to the standard ghost
number assignments in the associated BV formalism. From the point
of view of $\cH_M$ the standard ghost number assignment for the
fields, antifields and the antibracket are thus obtained by
shifting the ghost number by $\half$ so that the inner product
carries ghost number $-1$, while the antibracket is of ghost
number $1$.

To summarize, we have thus shown that~\eqref{ma2} is the proper
solution of the master equation for a first quantized BRST system
defined by $\hat H$ and $\hat \Omega$ on $\cH$. The antibracket is
determined by the inverse of the imaginary part of the inner
product $\langle \cdot,\cdot\rangle_M$ defined on $\cH_M$. Moreover, after
shifting, all physical fields are among the fields associated to
ghost number zero states while those associated to negative and
positive ghost number states are respectively ghost fields and
antifields.

\subsection{Time reparametrization invariant systems}\label{ris}

Suppose now that the BRST invariant Hamiltonian $H$ vanishes, as
in time reparametrization invariant systems. Suppose furthermore
that the original system has an odd inner product
$\langle \cdot,\cdot\rangle_{\rm st}$. This is the case for instance for the
relativistic particle or for the open bosonic string, where the
ghost pair $(\eta,{\cal P})$ associated to the mass-shell
constraint $p^2+m^2\approx 0$, respectively $L_0\approx 0$, is
quantized in the Schr\"odinger representation. According to our
discussion in subsection \ref{BRSTop}, in order to have an even
inner product and no fractionalization of the ghost number, the
system is extended to include the Lagrange multiplier $\lambda$
and its momentum $b$, together with the ghost pair $(\bar C,\rho)$
associated to the constraint $b\approx 0$. The BRST charge picks
up the additional term $b\rho$ and the pairs $(\lambda,b)$, $(\bar
C,\rho)$ are both quantized in the Schr\"odinger representation,
yielding the odd Hilbert space $\cH_{\lambda,\bar C}$. Hence, the
even Hilbert space $\cH$ with BRST charge $\hat \Omega$ is of the
form $\cH_{\rm st}\otimes_{\mathbb C}\cH_{\lambda,\bar C}$, where
$\cH_{\rm st}$ is odd with $\hat\Omega$ the tensor product of
$\hat \Omega_{\rm st}$ and $\hat b\hat\rho$.

The master action \eqref{ma2} can then be understood as resulting
from the original system described by the odd Hilbert space
$\cH_{\rm st}$, the BRST operator $\hat\Omega_{\rm st}$ and the
associated master action \bea {\mathbf S}_{\rm st}=\half
\langle \Psi_{\rm st},\hat\Omega_{\rm st}\Psi_{\rm st}\rangle_{\rm st},
 \eea
tensored with the system described by the even Hilbert space
$\cH_{\rm aux}= \cH_{\lambda,\bar C}\otimes_{\mathbb
C}\cH_{t,\theta}$ with inner product $\langle \cdot,\cdot\rangle_{\rm aux}$,
the BRST operator $\hat\Omega_{\rm aux}=\hat
b\hat\rho\otimes\mathbf 1+ \id\otimes\hat\theta\hat p_0$ and the
associated BRST charge \bea \mathbf \Omega_{\rm aux}=-\half
\langle \Psi_{\rm aux},\hat\Omega_{\rm aux}\Psi_{\rm aux}\rangle_{\rm
aux}\label{ma3}\,.\eea On the classical level, the auxiliary
system is described by the pairs $((\lambda,b),(t,p_0))$, the
constraints $b\approx 0\approx p_0$ and the ghost pairs $((\bar
C,\rho),(\theta,\pi))$. The associated BRST charge $\Omega_{\rm
aux}=\rho b+\theta p_0$ describes 2 trivial pairs and its
cohomology is generated by a constant. On the classical level, one
can thus simply get rid of these pairs. The question then is
whether first quantized BRST systems (and the associated classical
field theories) that differ by the quantization of classically
trivial pairs are equivalent.

If the 2 pairs had been quantized in the Fock instead of the
Schr\"odinger representation, then equivalence could have been
directly established. Indeed, all the states except for the Fock
vacuum $|\,0\rangle$ form quartets. Hence, according to the discussion
in subsection \ref{subtencon}, the proper master action
\eqref{ma3} can be consistently reduced to the master
action~${\mathbf S}_{\rm st}$.

\subsection{Quantization of trivial pairs and Chern-Simons}\label{CS}

When the 2 pairs are quantized in the Schr\"odinger
representation, it is convenient to rename them as
$\sigma^1,p_1,\sigma^2,p_2$, with $\commut{\hat\sigma^\alpha}{\hat
p_\beta}=i\delta^\alpha_\beta$. The associated fermionic ghost
pairs are $\eta^1,\cP_1,\eta^2,\cP_2$, with $\gh{\eta^\alpha}=1$
$\gh{\cP_\alpha}=-1$ and
$\commut{\hat\cP_\alpha}{\hat\eta^\beta}=-i\delta_\alpha^\beta$.
Wave functions and inner product are chosen as
\begin{multline}
\langle \phi,\psi\rangle=\int d\sigma^1 d\sigma^2 d\eta^1 d \cP_2\,
\bar\phi(\sigma,\eta^1,\cP_2)\psi(\sigma,\eta^1,\cP_2) \,=
\\
=\,\int d\sigma^1 d\sigma^2 d\eta^1 d \cP_2\,
h_{ij}\phi^i(\sigma,\eta^1,\cP_2)\psi^j(\sigma,\eta^1,\cP_2)\,,
\end{multline}
where in the second line we have expressed the hermitian inner
product in $\mathbb C$ in terms of two component real-valued wave
functions. The BRST charge and ghost number operators are given by
\begin{equation}
  \hat\Omega_{\rm aux}=-i\eta^1\dl{\sigma_1}-\dl{\sigma_2}\dl{\cP_2}\,,\qquad
\hat{\cal G}=\eta^1\dl{\eta^1}-\cP^2\dl{\cP^2}\,.
\end{equation}
In particular, states of the form $\psi(x)$, $\eta^1 \psi(x)$,
$\cP_2 \psi(x)$, and $\eta^1 \cP_2\psi(x)$ are respectively of
ghost degrees $0,1,-1$, and $0$.

The associated string field is
\begin{equation}
  \Psi=\int d^2\sigma\, |\sigma\rangle e_i\Big(\Phi^i_2(\sigma)
  +\eta^1P^i(\sigma)+\cP_2D^i(\sigma)+\eta^1\cP_2 \Phi^i_1(\sigma)\Big)\,,
\end{equation}
where the
$\Phi^i_2(\sigma),\Phi^i_1(\sigma),P^i(\sigma),D^i(\sigma)$ are
the coordinates on the supermanifold associated to the Hilbert
space. Their Grassmann parities and ghost numbers are
\begin{equation}
\begin{gathered}
  \p{\Phi^i_2}=\p{\Phi^i_1}=0,\qquad
  \p{P^i}=\p{D^i}=1\\
\gh{\Phi^i_2}=\gh{\Phi^i_1}=0\,, \quad \gh{P^i}=-1\,, \quad
\gh{D^i}=1\,,
\end{gathered}
\end{equation}
so that $\gh{\Psi}=0,\p{\Psi}=0$. From a geometrical point of
view, this supermanifold can be understood as the supermanifold of
maps from the supermanifold with coordinates
$\sigma^\alpha,\eta^1,\cP_2$ (configuration space) to  $\mathbb C$
viewed as a 2-dimensional real space. The Poisson bracket
corresponding to the symplectic form ${\rm Im}\langle \cdot,\cdot\rangle$ is
determined by
\begin{equation}
  \pb{\Phi^i_2(\sigma)}{\Phi^j_1(\sigma^\prime)}=-\omega^{ij}\delta(\sigma-\sigma^\prime)\,,\qquad
  \pb{D^i(\sigma)}{P^j(\sigma^\prime)}=\omega^{ij}\delta(\sigma-\sigma^\prime)\,.
\end{equation}

Let $J^i_j$ and $\omega_{ij}$ denote the complex structure and the
symplectic form on $\mathbb C$ respectively. Integrating out the
Grassmann odd variables $\eta^1,\cP_2$, using integrations by
parts and redefining the variables as $A_2^i=\Phi_2^i,A_1^i=-
J^i_j \Phi_1^j$ and $C^i=J^i_jD^j$, the BRST charge becomes
\begin{equation}
\label{eq:cs-charge}
  \mathbf \Omega_{\rm aux}
= -\int\,d\sigma^1 d\sigma^2\, g_{ij}\left[A_2^i\d_1 C^j -A_1^i
\d_2 C^j \right]\,.
\end{equation}
In terms of the new variables, the Poisson bracket is determined
by
\begin{equation}
\label{eq:fpb}
  \pb{A^i_1(\sigma)}{A^j_2(\sigma^\prime)}=-g^{ij}\delta(\sigma-\sigma^\prime)\,,\qquad
  \pb{C^i(\sigma)}{P^j(\sigma^\prime)}=-g^{ij}\delta(\sigma-\sigma^\prime)\,,
\end{equation}
where $g^{ij}=-J^i_k\omega^{kj}$ and satisfies
$g_{ij}g^{jk}=\delta^i_k$. The adjoint action $s=\pb{\cdot}{\mathbf
\Omega_{\rm aux}}$ reads
\begin{equation}
sA^i_\alpha=\d_\alpha C^i\,,\quad sC^i=0\,,\quad sP^i=-\d_1
A^i_2+\d_2 A^i_1\,.
\end{equation}
{}From this it follows that the BRST charge $\mathbf \Omega_{\rm
aux}$ is the BRST charge of complex Abelian Chern-Simons theory.
We conclude by giving some additional remarks on this BRST charge
and the associated master action.

\bigskip
\pagebreak

\noindent {\bf Remark 1: Superfield formulation}

\noindent Introducing the Grassmann odd superfields $\Lambda^i$ of
ghost number $1$
\begin{equation}
\label{eq:lambda} \Lambda^i(\sigma,\eta)=C^i(\sigma)+\eta^\alpha
A_\alpha^i(\sigma)+\eta^2\eta^1P^i(\sigma),
\end{equation}
the brackets \eqref{eq:fpb} are equivalent to
\begin{equation}
  \pb{\Lambda^i(\sigma,\eta)}{\Lambda^j(\sigma^\prime,\eta^\prime)}
=-g^{ij}\delta(\sigma-\sigma^\prime)\delta(\eta-\eta^\prime)\,.
\end{equation}
The BRST charge \eqref{eq:cs-charge} can then be rewritten as
\begin{equation}
\label{eq:cs-charge2}
  {\mathbf \Omega}_{\rm aux}=-\half \,\int d\sigma^1 d\sigma^2 d\eta^1 d\eta^2
  \,g_{ij}\Lambda^i (\eta^1 \d_1+\eta^2 \d_2)\Lambda^j\,.
\end{equation}
Applying the superfield reformulation to get the master action for
a BFV system described by the BRST charge $\mathbf \Omega$ and
vanishing Hamiltonian, one gets
\begin{equation}
  \mathbf S=\half\int d\sigma^0 d\sigma^1 d\sigma^2 d \eta^0 d\eta^1 d\eta^2 \,
g_{ij}\Big(\Lambda^i_S (\eta^0 \d_0+\eta^1 \d_1+\eta^2
\d_2)\Lambda^j_S\Big)\,,
\end{equation}
where the time coordinate is denoted by $\sigma^0$, the associated
Grassmann odd variable by $\eta^0$ and $\Lambda^i_S$ is the
superfield depending on $\sigma^\mu,\eta^\mu$, with $\mu=0,1,2$.

This coincides with the well-known AKSZ
representation~\cite{Alexandrov:1997kv} of the master action for
Abelian Chern Simons theory. The standard formulation can be
recovered by identifying $\eta^\mu$  with $d\sigma^\mu$ so that
the action takes the form $ \mathbf S=\half \int
g_{ij}(\Lambda^i_S \wedge d\Lambda^j_S)$, where $d$ is the de Rham
differential.

\bigskip

\noindent {\bf Remark 2: Coordinate representation for the ghosts}

\noindent If both ghost pairs are quantized in the coordinate
representation, one can arrive directly at~\eqref{eq:cs-charge2}
because the superfield \eqref{eq:lambda} then appears as the
projection of the string field on $\langle \sigma|$. Indeed, the BRST and
ghost number operator act on the states as
\begin{equation}
\hat\Omega=-i\eta^\alpha\dl{\sigma^\alpha}\,,\qquad \hat{\cal
G}=\eta^\alpha \dl{\eta_\alpha}-1\,.
\end{equation}
In particular, states of the form $\psi(\sigma)$, $\eta^\alpha
\psi_\alpha(\sigma)$, and $\eta^2\eta^1 \chi(\sigma)$ are
respectively of ghost degrees $-1,0$, and $1$. In order to have a
bosonic field theory, we assign Grassmann parity $k~{\rm mod}~2$
to the states of ghost degree $k$. This implies defining the inner
product on the super Hilbert space by
\begin{equation}
\label{inner} \langle \phi,\psi\rangle=-i(-1)^{\p{\psi}}\int d\sigma^1
d\sigma^2 d\eta^1 d\eta^2\,\,
\overline{\phi(\sigma,\eta)}\,\psi(\sigma,\eta)\,,
\end{equation}
so that
$\overline{\langle \phi,\psi\rangle}=(-1)^{\p{\phi}\p{\psi}}\langle \psi,\phi\rangle$. The
associated string field of total ghost number and Grassmann parity
zero is now
\begin{equation}
  \Psi=\int d^2\sigma\, |\sigma\rangle e_i\Lambda^i(\sigma,\eta)\,,
\end{equation}
and the corresponding ${\mathbf
\Omega}=-\half\langle \Psi,\hat\Omega\Psi\rangle$ coincides
with~\eqref{eq:cs-charge2}.

\bigskip

\noindent {\bf Remark 3: Non Abelian Chern-Simons theory}

\noindent In ~\cite{Alexandrov:1997kv}, the expression for the
master action is derived for the Lie algebra of a compact Lie
group. Using the same reasoning as above, it can easily be shown
that the associated BRST charge can be compactly written as \bea
\mathbf \Omega = -\half \,\int d\sigma^1 d\sigma^2 d\eta^1
d\eta^2\Big(
  \,g_{IJ}\Lambda^I (\eta^1 \d_1+\eta^2 \d_2)\Lambda^J+
  \frac{1}{3}f_{IJK}\Lambda^I\Lambda^J\Lambda^K\Big)\,,
\eea where $g_{IJ}$ denotes the invariant metric and
$f_{IJK}=g_{IL}f^L_{JK}$. Put differently, the BRST charge for
Chern-Simons theory has exactly the same form as the AKSZ master
action (and therefore the classical action). Only the source
supermanifolds are different: for the master action, the
superdimension is $(3|3)$, while for the BRST charge it is
$(2|2)$.

Similar remarks apply for the BRST charge and the master action of
the Poisson sigma model
\cite{Cattaneo:1999fm,Batalin:2001fh,Cattaneo:2001ys}.

\section{Discussion}

The new feature of the present paper is the shift of emphasis, on
the level appropriate for second quantization, from the master
action to the BRST charge. Given a gauge system quantized
according to the Hamiltonian BRST approach, one can always make
the number of constraints even if necessary. The associated object
$\mathbf \Omega=-\half\langle \Psi,\hat\Omega\Psi\rangle$ is then a nilpotent
BRST charge with respect to the even Poisson bracket induced by
the imaginary part of the inner product. Out of this BRST charge,
the master action can be constructed according to a standard
procedure. In particular, for closed string field theory for
instance, $\mathbf\Omega$ is naturally a BRST charge, without the
necessity of adding trivial pairs. We plan to discuss this issue
in more details in future work.

\section*{Acknowledgements}
\addcontentsline{toc}{section}{Acknowledgments}

G.B.~wants to thank M.~Henneaux for suggesting the problem. Useful
discussions with I.~Batalin, G.~Bonelli, L.~Houart,
A.~Semikhatov and I.~Tipunin are gratefully acknowledged. The work
of GB and MG is supported in part by the ``Actions de Recherche
Concert{\'e}es'' of the ``Direction de la Recherche
Scientifique-Communaut{\'e} Fran\c{c}aise de Belgique", by a
``P{\^o}le d'Attraction Interuniversitaire'' (Belgium), by
IISN-Belgium, convention 4.4505.86, by the INTAS grant 00-00262,
and by the European Commission RTN program HPRN-CT00131, in which
the authors are associated to K.~U.~Leuven. GB is also supported
by Proyectos FONDECYT 1970151 and 7960001 (Chile), while MG is
supported by RFBR grants 02-01-00930 and 02-01-06096.

\subsection*{Appendix A: Formulation of supermanifold in terms of complex coordinates}
\addcontentsline{toc}{subsection}{Appendix A: Formulation of
supermanifold in terms of complex coordinates}

The geometrical structures on the supermanifold $\manM_\cH$ can be
conveniently expressed in terms of complex coordinates. We follow
\cite{KobNom2}.

Consider the complexification $\cH^{\mathbb C}=\cH\tensor_{\mathbb
R} {\mathbb C}$ where $\cH$ is considered as above as a superspace
over $\mathbb R$. The complex conjugation of a vector of the form
$\alpha\psi,\,\,\psi \in \cH,\,\, \alpha \in {\mathbb C}$ is
defined as $\bar\alpha \psi$ so that the original Hilbert space
$\cH$ is a subspace (over $\mathbb R$) of vectors satisfying
$\bar\psi=\psi$.

All $\mathbb R$-linear operations on $\cH$ can be be extended to
$\cH^{\mathbb C}$ by $\mathbb C$-linearity. In particular,
$\cH^{\mathbb C}$ decomposes as $\cH^{\mathbb
C}=\cH^{1,0}\oplus\cH^{0,1}$ with
\begin{equation}
  \hat J\psi=i \psi \quad\forall \psi\in\cH^{1,0}\,,
  \qquad  \hat J\phi=-i\phi \quad\forall \phi\in \cH^{0,1}\,.
\end{equation}
Complex conjugation defines a real linear isomorphism between
$\cH^{1,0}$ and $\cH^{0,1}$. Introducing a basis $\{e_\alpha\}$
for $\cH^{1,0}$ and  $\{e_{\bar \alpha}\}$ for $\cH^{0,1}$ such
that $\overline{e_\alpha}=e_{\bar\alpha}$, the inner product
$\inner{\cdot}{\cdot}$ extended by $\mathbb C$ bi-linearity is determined by
\begin{equation}
\langle e_{\bar\alpha},e_\beta\rangle=(-1)^{\p{\bar\alpha}}h_{\bar\alpha\beta}\,,\qquad
\langle e_\alpha,e_{\bar\beta}\rangle=\langle e_\alpha,e_{\beta}\rangle=\langle e_{\bar\alpha},e_{\bar\beta}\rangle=0\,.
\end{equation}
The graded-symmetric and graded-antisimmetric components $g$ and
$\omega$ of $\langle ,\rangle$ are determined by
\begin{align}
g(e_{\bar\alpha},e_\beta)=&\,\half
(-1)^{\p{\bar\alpha}}h_{\bar\alpha\beta}\,, \qquad
g(e_{\alpha},e_{\bar\beta})=\half (-1)^{\p{\bar\beta}
+\p{\alpha}\p{\bar\beta}}h_{\bar\beta\alpha}\,,\\
\omega(e_{\bar\alpha},e_\beta)=&\,\frac{1}{2i}
(-1)^{\p{\bar\alpha}}h_{\bar\alpha\beta}\,, \qquad
\omega(e_{\alpha},e_{\bar\beta})=-\frac{1}{2i}
(-1)^{\p{\bar\beta}+\p{\alpha}\p{\bar\beta}}h_{\bar\beta\alpha}\,,
\end{align}
with all other components vanishing. From
$\langle {e}_{\bar\alpha},{e}_\beta\rangle=\langle e_\alpha+e_{\bar\alpha},e_\beta+e_{\bar\beta}\rangle$
and  the fact that $e_\alpha+e_{\bar\alpha}$ is a real vector it
follows that
\begin{equation}
  \overline{h_{\bar\alpha\beta}}=(-1)^{1+(1+\p{\bar\alpha})(1+\p{\beta})}h_{\bar\beta\alpha}\,.
\end{equation}
By considering $\cH$ as a space over $\mathbb R$, $\mathbb
C$-linear operators are identified with $\mathbb R$ linear
operators commuting with $\hat J$. In the basis $e_\alpha,e_{\bar
\alpha}$, this means that the matrix of such an operator $\hat A$
is block-diagonal with only the diagonal blocks $A^\alpha_\beta$
and $A^{\bar \alpha}_{\bar \beta}$ nonvanishing. The fact that
$\hat A$ is extended from $\cH$ to $\cH^{\mathbb C}$ by $\mathbb
C$-linearity implies that $\hat A$ maps real vectors to real ones
so that $\hat A(e_\alpha+e_{\bar \alpha})$ is again a real vector,
which in turn implies that $\overline{A^\alpha_\beta}=A^{\bar
\alpha}_{\bar \beta}$.

Associated to the basis elements $e_{\alpha},e_{\bar\beta}$, one
then introduces variables $\Psi^\alpha,\Psi^{\bar\alpha}$ with
$\p{\Psi^\alpha}=\p{\Psi^{\bar\alpha}}=\p{\alpha}$ and
$\gh{\Psi^\alpha}=\gh{\Psi^{\bar\alpha}}=-\gh{e_\alpha}$ and
considers $\mathfrak G^{\mathbb C}$, the algebra of complex valued
functions in these variables. The complex conjugation in
$\cH^{\mathbb C}$ naturally determines a complex conjugation in
$\mathfrak G^{\mathbb C}$ through
$\overline{\Psi^\alpha}=\Psi^{\bar\alpha}$ so that the real
elements of $\mathfrak G^{\mathbb C}$ can then be identified with
$\mathfrak G$. The symplectic form $\omega$ on $\cH^{\mathbb C}$
determines a Poisson bracket in $\mathfrak G^{\mathbb C}$
determined by
\begin{equation}
\{\Psi^\alpha,\Psi^{\bar\beta}\}=2i
h^{\alpha\bar\beta}\,,\qquad
h_{\bar\alpha\beta}h^{\beta\bar\gamma}=\delta^{\bar\alpha}_{\bar\gamma}\,.
\end{equation}
The string field is then given by $\Psi=\Psi^\alpha
e_\alpha+\Psi^{\bar\alpha}e_{\bar\alpha}$.

\subsection*{Appendix B: Dirac observables and cohomology of $\mathbf {\langle \Psi,\hat\Omega\Psi\rangle}$}
\addcontentsline{toc}{subsection}{Appendix B: Dirac observables
and cohomology of $\langle \Psi,\hat\Omega\Psi\rangle$}

\def\theequation{B.\arabic{equation}}
\setcounter{equation}{0} Formally, one can assume that a real basis
$\{e_a\}\equiv \{e_i,f_m,g_m\}$ in the Hilbert space $\cH$ is
chosen such that \bea (-\hat J\hat \Omega) e_i=0\,,\quad(-\hat
J\hat \Omega) f_m=g_m\,,\quad (-\hat J \hat \Omega) g_m=0\,.\eea
The associated coordinates of the supermanifold are
$\{\Psi^i\}\equiv\{\Xi^i,\Upsilon^m,\Phi^m\}$. On the one hand,
the differential $\{\cdot,\mathbf \Omega\}$ becomes \bea
\{\cdot,\mathbf \Omega\}=\ddr{\cdot}{\Phi^m}\Upsilon^m\,,\eea so that
$H(\{\cdot,\mathbf \Omega\})$ is isomorphic to the algebra of
functions in $\Xi^i$ alone. On the other hand, the constraints are
given by $\Upsilon^m\approx 0$. Antihermiticity and nilpotency of
$-\hat J\hat \Omega$ implies that the symplectic structure in the
basis $\{e_i,f_m,g_m\}$ becomes
\begin{equation}
\begin{pmatrix}
  \omega_{ij} & \omega_{im} & 0 \\
  \omega_{kj} & \omega^\prime_{km} & \omega_{kn} \\
  0 & \omega_{lm} & 0
\end{pmatrix}\,,
\end{equation}
with both $\omega_{ij}$ and $\omega_{lm}$ non degenerate. The
inverse has the form
\begin{equation}
\begin{pmatrix}
\omega^{ji} & 0& \tilde\omega^{ni} \\
0 & 0 & \omega^{nr} \\
\tilde\omega^{jp}& \omega^{mp} & \tilde\omega^{np}
\end{pmatrix}\,,
\end{equation}
with both $\omega^{ji}$ and $\omega^{mp}$ non degenerate. This
implies that the adjoint action in the Poisson bracket of the
constraints $\Upsilon^m$ generate shifts in the $\Phi^m$, which
are thus coordinates along the gauge orbits. Hence, equivalence
classes of Dirac observables also correspond to functions in
$\Xi^i$ alone.

\subsection*{Appendix C: Quantum BRST state cohomology of tensor products}
\addcontentsline{toc}{subsection}{Appendix C: Quantum BRST state
cohomology of tensor products}

\def\theequation{C.\arabic{equation}}
\setcounter{equation}{0}

Let $\{k_\alpha\}$ and $\{K_\Lambda\}$ be bases over $\mathbb C$
in $\cH_1$ and $\cH_2$ respectively. Then the vectors
$k_{\alpha\Lambda}=k_\alpha \densor K_\Lambda$ provide a basis (over ${\mathbb C}$) of
the tensor product $\cH_1\densor_{\mathbb C} \cH_2$. Assume that in the bases
$k_\alpha=\{k_\theta,f_\gamma,g_\gamma\}$ and
$K_\Lambda=\{K_\Theta,F_\Gamma,G_\Gamma\}$, the BRST charges
$\hat\Omega_1$ and $\hat\Omega_2$ take the Jordan form,
\begin{equation}
  \begin{aligned}
        \hat\Omega_1 k_\theta&=&0\,, \quad \hat\Omega_1 f_m&=&g_m\,, \quad \hat\Omega_1 g_m&=&0\,,\\
        \hat\Omega_2 K_\Theta&=&0\,, \quad \hat\Omega_2 F_M&=&G_M\,, \quad \hat\Omega_2 G_M&=&0\,.
\end{aligned}
\end{equation}
One then can check that the vectors
\begin{equation}
\begin{gathered}
  k_{\theta\Theta}=k_\theta\densor K_\Theta\\
f_{\gamma\Gamma}=f_\gamma\densor F_\Gamma\,,\quad
{\tilde f}_{\gamma \Gamma}=\half(g_\gamma\densor F_\Gamma-(-1)^{\p{f_\gamma}}f_\gamma\densor G_\Gamma)\,,\\
f^0_{\theta\Gamma}=k_\theta\densor F_\Gamma\,,\quad
f^0_{\gamma \Theta}=f_\gamma\densor K_\Theta\\
g_{\gamma\Gamma}=g_\gamma\densor
F_\Gamma+(-1)^{\p{f_\gamma}}f_\gamma\densor G_\Gamma\,,\quad
{\tilde g}_{\gamma\Gamma}=(-1)^{\p{g_\gamma}}g_\gamma\densor G_\Gamma\,,\\
g^0_{\theta \Gamma}=(-1)^{\p{k_\theta}}k_\theta\densor
G_\Gamma\,,\quad
g^0_{\gamma \Theta}=g_\gamma\densor K_\Theta\\
\end{gathered}
\end{equation}
form a Jordan basis for $\hat\Omega_T$. Hence, the cohomology of
$\hat\Omega_T$ is the linear span over $\mathbb C$ of
$k_{\theta\Theta}=k_\theta\densor K_\Theta$, so that
$H(\hat\Omega_T,\cH_T)=H(\hat\Omega_1,\cH_1){\densor}_{\mathbb C}
H(\hat\Omega_2,\cH_2)$.

\addcontentsline{toc}{section}{Bibliography}

\providecommand{\href}[2]{#2}\begingroup\raggedright\endgroup


\begin{thebibliography}{10}

\bibitem{Thorn:1987qj}
C.~B. Thorn, ``Perturbation theory for quantized string fields,''
{\em Nucl.
  Phys.} {\bf B287} (1987)
61.

\bibitem{Bochicchio:1987zj}
M.~Bochicchio, ``Gauge fixing for the field theory of the bosonic
string,''
  {\em Phys. Lett.} {\bf B193} (1987)
31.

\bibitem{Bochicchio:1987bd}
M.~Bochicchio, ``String field theory in the {S}iegel gauge,'' {\em
Phys. Lett.}
  {\bf B188} (1987)
330.

\bibitem{Thorn:1989hm}
C.~B. Thorn, ``String field theory,'' {\em Phys. Rept.} {\bf 175}
(1989) 1--101.

\bibitem{Siegel:1988yz}
W.~Siegel, \textit{Introduction to string field theory,} { World
Scientific}
  (1988),
\href{http://www.arXiv.org/abs/hep-th/0107094}{{\tt
hep-th/0107094}}.

\bibitem{Witten:1986cc}
E.~Witten, ``Noncommutative geometry and string field theory,''
{\em Nucl.
  Phys.} {\bf B268} (1986)
253.

\bibitem{Batalin:1981jr}
I.~A. Batalin and G.~A. Vilkovisky, ``Gauge algebra and
quantization,'' {\em
  Phys. Lett.} {\bf B102} (1981)
27--31.

\bibitem{Batalin:1983wj}
I.~A. Batalin and G.~A. Vilkovisky, ``Feynman rules for reducible
gauge
  theories,'' {\em Phys. Lett.} {\bf B120} (1983)
166--170.

\bibitem{Batalin:1983jr}
I.~A. Batalin and G.~A. Vilkovisky, ``Quantization of gauge
theories with
  linearly dependent generators,'' {\em Phys. Rev.} {\bf D28} (1983)
2567--2582.

\bibitem{Batalin:1984ss}
I.~A. Batalin and G.~A. Vilkovisky, ``Closure of the gauge
algebra, generalized
  {L}ie equations and {F}eynman rules,'' {\em Nucl. Phys.} {\bf B234} (1984)
106--124.

\bibitem{Batalin:1985qj}
I.~A. Batalin and G.~A. Vilkovisky, ``Existence theorem for gauge
algebra,''
  {\em J. Math. Phys.} {\bf 26} (1985)
172--184.

\bibitem{Henneaux:1992ig}
M.~Henneaux and C.~Teitelboim, {\em Quantization of {G}auge
{S}ystems}.
\newblock Princeton University Press, 1992.

\bibitem{Gomis:1995he}
J.~Gomis, J.~Par\'{\i}s, and S.~Samuel, ``Antibracket, antifields
and gauge
  theory quantization,'' {\em Phys. Rept.} {\bf 259} (1995) 1--145,
\href{http://www.arXiv.org/abs/hep-th/9412228}{{\tt
hep-th/9412228}}.

\bibitem{Fradkin:1975cq}
E.~S. Fradkin and G.~A. Vilkovisky, ``Quantization of relativistic
systems with
  constraints,'' {\em Phys. Lett.} {\bf B55} (1975)
224.

\bibitem{Batalin:1977pb}
I.~A. Batalin and G.~A. Vilkovisky, ``Relativistic {S} matrix of
dynamical
  systems with boson and fermion constraints,'' {\em Phys. Lett.} {\bf B69}
  (1977)
309--312.

\bibitem{Fradkin:1978xi}
E.~S. Fradkin and T.~E. Fradkina, ``Quantization of relativistic
systems with
  boson and fermion first and second class constraints,'' {\em Phys. Lett.}
  {\bf B72} (1978)
343.

\bibitem{Henneaux:1985kr}
M.~Henneaux, ``Hamiltonian form of the path integral for theories
with a gauge
  freedom,'' {\em Phys. Rept.} {\bf 126} (1985) 1.

\bibitem{Kibble:1979tm}
T.~W.~B. Kibble, ``Geometrization of quantum mechanics,'' {\em
Commun. Math.
  Phys.} {\bf 65} (1979)
189.

\bibitem{Heslot:1985xx}
A.~Heslot, ``Quantum mechanics as a classical theory,'' {\em Phys.
Rev. D} {\bf
  31} (1985) 1341--1348.

\bibitem{Hatfield:1992rz}
B.~Hatfield, ``Quantum field theory of point particles and
strings,''. Redwood
  City, USA: Addison-Wesley (1992) 734 p. (Frontiers in physics, 75).

\bibitem{Schilling:1996xx}
T.~Schilling, {\em Geometry of quantum mechanics}.
\newblock PhD thesis, The Pennsylvania State University, 1996.

\bibitem{Ashtekar:1997ud}
A.~Ashtekar and T.~A. Schilling, ``Geometrical formulation of
quantum
  mechanics,''
\href{http://www.arXiv.org/abs/gr-qc/9706069}{{\tt
gr-qc/9706069}}.

\bibitem{Zwiebach:1993ie}
B.~Zwiebach, ``Closed string field theory: Quantum action and the
{B}-{V}
  master equation,'' {\em Nucl. Phys.} {\bf B390} (1993) 33--152,
\href{http://www.arXiv.org/abs/hep-th/9206084}{{\tt
hep-th/9206084}}.

\bibitem{Grigoriev:2000zg}
M.~A. Grigoriev, A.~M. Semikhatov, and I.~Y. Tipunin, ``{BRST}
formalism and
  zero locus reduction,'' {\em J. Math. Phys.} {\bf 42} (2001) 3315--3333,
\href{http://www.arXiv.org/abs/hep-th/0001081}{{\tt
hep-th/0001081}}.

\bibitem{Batalin88}
I.~Batalin and E.~Fradkin, ``{O}peratorial quantization of
dynamical systems
  subject to constraints. {A} further study of the construction,'' {\em Ann.
  Inst. Henri Poincare (Phys. Theor.)} {\bf 49} (1988) 145--214.

\bibitem{Siegel:1989nh}
W.~Siegel, ``Batalin-{V}ilkovisky from {H}amiltonian {BRST},''
{\em Int. J.
  Mod. Phys.} {\bf A4} (1989)
3951.

\bibitem{Batlle:1989if}
C.~Batlle, J.~Gomis, J.~Par\'{\i}s, and J.~Roca, ``Lagrangian and
{H}amiltonian
  {BRST} formalisms,'' {\em Phys. Lett.} {\bf B224} (1989)
288.

\bibitem{Fisch:1989rm}
J.~M.~L. Fisch and M.~Henneaux, ``Antibracket - antifield
formalism for
  constrained {H}amiltonian systems,'' {\em Phys. Lett.} {\bf B226} (1989) 80.

\bibitem{Henneaux:1990ua}
M.~Henneaux, ``Elimination of the auxiliary fields in the
antifield
  formalism,'' {\em Phys. Lett.} {\bf B238} (1990) 299.

\bibitem{Batlle:1990zq}
C.~Batlle, J.~Gomis, J.~Par\'{\i}s, and J.~Roca, ``Field -
antifield formalism
  and {H}amiltonian {BRST} approach,'' {\em Nucl. Phys.} {\bf B329} (1990)
  139--154.

\bibitem{Dresse:1990dj}
A.~Dresse, P.~Gr\'egoire, and M.~Henneaux, ``Path integral
equivalence between
  the extended and nonextended {H}amiltonian formalisms,'' {\em Phys. Lett.}
  {\bf B245} (1990) 192.

\bibitem{Dresse:1991ba}
A.~Dresse, J.~M.~L. Fisch, P.~Gr\'egoire, and M.~Henneaux,
``Equivalence of the
  {H}amiltonian and {L}agrangian path integrals for gauge theories,'' {\em
  Nucl. Phys.} {\bf B354} (1991) 191--217.

\bibitem{Grigorian:1991zs}
G.~V. Grigorian, R.~P. Grigorian, and I.~V. Tyutin, ``Equivalence
of
  {L}agrangian and {H}amiltonian {BRST} quantization. {S}ystems with
  first-class constraints,'' {\em Sov. J. Nucl. Phys.} {\bf 53} (1991)
1058--1061.

\bibitem{Grigoriev:1999qz}
M.~A. Grigoriev and P.~H. Damgaard, ``Superfield {BRST} charge and
the master
  action,'' {\em Phys. Lett.} {\bf B474} (2000) 323--330,
\href{http://www.arXiv.org/abs/hep-th/9911092}{{\tt
hep-th/9911092}}.

\bibitem{Siegel:1989ip}
W.~Siegel, ``Relation between {B}atalin-{V}ilkovisky and first
quantized style
  {BRST},'' {\em Int. J. Mod. Phys.} {\bf A4} (1989) 3705.

\bibitem{Dayi:1993fk}
O.~F. Dayi, ``A general solution of the {BV} master equation and
{BRST} field
  theories,'' {\em Mod. Phys. Lett.} {\bf A8} (1993) 2087--2098,
\href{http://www.arXiv.org/abs/hep-th/9305038}{{\tt
hep-th/9305038}}.

\bibitem{Siegel:1991zf}
W.~Siegel, ``Boundary conditions in first quantization,'' {\em
Int. J. Mod.
  Phys.} {\bf A6} (1991)
3997--4008.

\bibitem{Barnich:1996mr}
G.~Barnich and M.~Henneaux, ``Isomorphisms between the
{B}atalin-{V}ilkovisky
  antibracket and the {P}oisson bracket,'' {\em J. Math. Phys.} {\bf 37} (1996)
  5273--5296,
\href{http://www.arXiv.org/abs/hep-th/9601124}{{\tt
hep-th/9601124}}.

\bibitem{DeligneFreed1}
P.~Deligne and W.~Morgan, {\em Quantum {F}ields and {S}trings: {A}
{C}ourse for
  {M}athematicians, {P}art {I}}, ch.~Notes on {S}upersymmetry.
\newblock American Mathematical Society, 1999.

\bibitem{Gaberdiel:1997ia}
M.~R. Gaberdiel and B.~Zwiebach, ``Tensor constructions of open
string theories
  {I}: Foundations,'' {\em Nucl. Phys.} {\bf B505} (1997) 569--624,
\href{http://www.arXiv.org/abs/hep-th/9705038}{{\tt
hep-th/9705038}}.

\bibitem{Sen:1994ic}
A.~Sen and B.~Zwiebach, ``A note on gauge transformations in
  {B}atalin-{V}ilkovisky theory,'' {\em Phys. Lett.} {\bf B320} (1994) 29--35,
\href{http://www.arXiv.org/abs/hep-th/9309027}{{\tt
hep-th/9309027}}.

\bibitem{Grigoriev:1998gn}
M.~A. Grigoriev, A.~M. Semikhatov, and I.~Y. Tipunin, ``Gauge
symmetries of the
  master action in the {B}atalin- {V}ilkovisky formalism,'' {\em J. Math.
  Phys.} {\bf 40} (1999) 1792--1806,
\href{http://www.arXiv.org/abs/hep-th/9804156}{{\tt
hep-th/9804156}}.

\bibitem{Alexandrov:1997kv}
M.~Alexandrov, M.~Kontsevich, A.~Schwartz, and O.~Zaboronsky,
``The geometry of
  the master equation and topological quantum field theory,'' {\em Int. J. Mod.
  Phys.} {\bf A12} (1997) 1405--1430,
\href{http://www.arXiv.org/abs/hep-th/9502010}{{\tt
hep-th/9502010}}.

\bibitem{Cattaneo:1999fm}
A.~S. Cattaneo and G.~Felder, ``A path integral approach to the
{K}ontsevich
  quantization formula,'' {\em Commun. Math. Phys.} {\bf 212} (2000) 591--611,
\href{http://www.arXiv.org/abs/math.qa/9902090}{{\tt
math.qa/9902090}}.

\bibitem{Batalin:2001fh}
I.~Batalin and R.~Marnelius, ``Generalized {P}oisson sigma
models,'' {\em Phys.
  Lett.} {\bf B512} (2001) 225--229,
\href{http://www.arXiv.org/abs/hep-th/0105190}{{\tt
hep-th/0105190}}.

\bibitem{Cattaneo:2001ys}
A.~S. Cattaneo and G.~Felder, ``On the {AKSZ} formulation of the
{P}oisson
  sigma model,'' {\em Lett. Math. Phys.} {\bf 56} (2001) 163--179,
\href{http://www.arXiv.org/abs/math.qa/0102108}{{\tt
math.qa/0102108}}.

\bibitem{KobNom2}
S.~Kobayashi and K.~Nomizu, {\em Foundations of differential
geometry},
  vol.~II.
\newblock John Wiley \& Sons, 1996~ed., 1969.

\end{thebibliography}
\end{document}